# Quasi-particles and their absence in photoemission spectroscopy

*(American Physical Society 2002 Isakson Prize Paper
written for Solid State Communications, the supporter of the prize)*

J.W. Allen

Randall Laboratory, University of Michigan, Ann Arbor, MI 48109-1120 USA

Abstract

The elucidation of Landau Fermi liquid quasi-particles and their absence in strongly correlated electron systems lies at the heart of modern research on the quantum mechanics of electrons in condensed matter. Photoemission spectroscopy of the single particle spectral function is a central experimental tool for such studies. A general paradigm of quasi-particle formation is the Fermi level resonance associated with the Kondo physics of the Anderson impurity model, as shown by the formal appearance of an effective impurity problem in dynamic mean field theories of general lattice systems which may or may not literally display Kondo physics. A general paradigm of quasi-particle absence is the Luttinger liquid physics of the Tomonaga-Luttinger model. This paper presents an overview of the theoretical ideas and shows examples in photoemission spectra of quasi-particle formation in impurity and lattice Kondo and non-Kondo correlated electron systems, contrasted with quasi-particle absence in a quasi-one dimensional system.

1. Introduction

In the last quarter century photoemission spectroscopy (PES) [1] has developed steadily to take its place with optical and Raman spectroscopy and with neutron scattering as an established general technique for studying the electronic structure of condensed matter systems. PES differs from these other spectroscopies in that they measure two particle correlation functions whereas, under the assumptions that the photoemission event can be described within the sudden approximation and that the photoemission matrix element can be adequately disentangled from the measured spectrum, PES is the method whereby the spectral function for single-particle electron removal can be measured [2]. Angle resolved PES (ARPES) [3] and PES yield the **k**-resolved and **k**-summed spectral functions, respectively, where **k** is the crystal momentum. The partner technique of inverse photoemission spectroscopy (IPES), also known as Bremsstrahlung isochromat spectroscopy (BIS) when performed with x-rays [4], measures the single particle spectral function to add an electron and so the combination of the two can yield the total single particle spectral function [5] $\rho(\mathbf{k},\omega) = -(1/\pi)$ Im $G(\mathbf{k},\omega)$ where G is the many-body single particle Green's function and $\omega$ is frequency or energy. Both techniques, but in recent years especially PES because it has developed to have much better energy resolution, have had a strong impact on the study of materials in the class of strongly correlated electron systems (SCES) [6]. This article presents examples of such work in the context of the elucidation of Landau Fermi liquid (FL) quasi-particles and their absence in SCES, a subject that lies at the heart of modern research on the quantum mechanics of electrons in condensed matter.

The Fermi liquid picture put forth by Landau [7] requires a non-interacting reference system that has the same ground state symmetry as that of the interacting system of interest. It is then supposed that as the interaction is turned on smoothly the low lying single-particle excitations retain their quantum numbers and evolve on a one-for-one basis into the "quasi-particles" of the interacting system. P.W. Anderson [8] has discussed this picture under the term of "adiabatic continuity." Landau applied the idea to a Fermi system, liquid helium, having translational symmetry so that momentum is a good quantum number. Important implications of the basics of the idea are then that the Fermi surface (FS) is robust, that its volume is preserved in the interacting system, and that many-body effects are manifested only by renormalizations of parameters such as Fermi surface masses. These are the central principles of the standard theory of electrons in metals and indeed the term "Fermi liquid" is often taken as synonymous with "normal metal." One testimony to the strength of the Fermi liquid concept is that many "heavy Fermion" metals, with interactions strong enough to produce effective masses as large as 1000, nonetheless appear to be Fermi liquids. [9,10]

The generality of the Fermi liquid idea transcends its well-known applications to metals. For example, although such systems are not discussed in this article, one could say that normal semiconductors like Si are also Fermi liquids in the sense of adiabatic continuity, but with renormalized gaps [11] and band edge masses. Nozieres [12] has applied the idea to Kondo impurity physics, the phenomenon in which an impurity magnetic moment coupled to an otherwise non-interacting electron gas is quenched for temperatures T below the Kondo temperature $T_K$ because the ground state of the entire system is a singlet. Nozieres showed that an electron gas with a non-magnetic charged impurity forms a suitable reference for this impurity FL. Here the effect of interactions is manifested as an enhanced impurity contribution to the specific heat and magnetic susceptibility. It is precisely the widespread success and apparent generality of FL ideas that make the study of their failure so important and interesting. A number of theoretical non-Fermi liquid (NFL) scenarios are now known, two of which are the "Luttinger" physics [13] of the interacting one-dimensional electron gas, discussed further below, and the strange behavior of multi-channel impurity Kondo systems [14,15]. In both cases non-interacting reference systems do not exist. Thus metallic NFL behavior is a part of the current broad theoretical emphasis, cutting across several sub-fields of physics, on non-perturbative field theories [16].

In the broadest terms, quasi-particle physics for an SCES material entails in the single-particle spectrum $\rho(\mathbf{k},\omega)$ a division of the spectral weight between a more or less sharp peak near the Fermi energy $E_F$ (or for a semiconductor near the top and bottom of the valence and conduction band edges, respectively) and more or less broad features which can be, but need not be, mostly at higher energies. The former can be traced back to the non-interacting reference state and so is often called the "coherent part" or "quasi-particle peak." The latter is often called the "incoherent part" or "incoherent background." The division of weight between the two parts is variable, with less and less weight in the coherent part as the system becomes more and more strongly correlated. The weight in the quasi-particle peak is linked to single-particle transport properties such as the specific heat. Although this general scenario plays out in a variety of particular ways in the spectra of various systems there are also commonalities imposed by Fermi surface and Fermi energy sum rules. This article aims at giving some idea of both the diversity and the commonalities by presenting spectra from several different systems studied by the author and his collaborators.

At the simplest level, the spectra of NFL systems do not display a quasi-particle peak, but are entirely incoherent. Nonetheless the incoherent part can have identifying structure that is peculiar to the particular case at hand and this article presents an example of such behavior also. More detailed accounts of many of these topics are contained in two papers [17,18] that are part of a recent collection of articles [19] focusing on many aspects and applications of photoemission for SCES materials.

2. Experimental aspects

Detailed expositions of photoemission spectroscopy are available in books and articles [1,2,20,21]. The following is a brief summary of background information useful for the material of this article.

2.1 ARPES basics

The basic experiment consists of measuring the angle and kinetic energy dependences of the intensity of photoelectrons emitted from a single crystal sample. As the electron detector angles are varied for fixed photon energy, a spherical **k**-space surface is traversed. Tuning the photon energy changes the radius of the spherical surface. Although the component of **k** parallel to the sample surface is conserved in the photoemission process, the perpendicular component $k_\perp$ changes because the photoelectron traverses the surface potential. Relating the value of $k_\perp$ measured outside the sample to its value inside the sample requires modeling the surface potential, e.g. by an "inner potential step" $V_0$ [20]. Determining $V_0$ and thus being accurately oriented in **k**-space entails measuring over a wide photon energy range to find repeating spectral patterns that can be registered correctly with repeating Brillouin zones for a suitable value of $V_0$. This process and indeed the spectra themselves can be compromised by the photoelectron lifetime, which causes $k_\perp$ broadening. The photoelectron lifetime also makes a contribution to the photohole linewidth that is additional to that of the intrinsic spectral function one seeks to measure. This contribution is proportional to the perpendicular photohole velocity [2, 21]. All these basic aspects of ARPES greatly favor the study of quasi-low dimensional systems.

Small electron elastic escape depths [22] cause the technique to be surface sensitive. This fact limits the observation time to the period over which a prepared surface remains free of contaminants, which necessitates excellent vacuum. Even for a clean surface there is the need to distinguish electronic structure that is peculiar to the surface and not characteristic of the bulk. This need is met by testing a spectral feature for extreme sensitivity to deliberate surface contamination or by decreasing the effective escape depth by increasing the detection angle away from the surface normal. The escape depth also varies with kinetic energy, being nominally a U-shaped curve with a minimum somewhere in a broad range around 100 eV. Bulk sensitivity is thus increased by using a high photon energy [23], but at the price of generally reduced photoemission cross-sections, of decreased **k**-resolution for a given detector angle resolution, and until very recently, of reduced photon energy resolution.

One sees from the discussion of this sub-section that the basic experimental variables of detector angle and photon energy are overworked. Varying the photon energy simultaneously changes the perpendicular k-component, the photoemission cross-section and the bulk sensitivity, while varying the detector angle changes the parallel k-components and the bulk sensitivity. It is a matter of both skill and luck to choose a material for detailed study in which these various aspects can be adequately disentangled.

2.2 ARPES techniques

It is often desirable to distinguish the character of states, e.g., their angular momentum or their association with certain atoms. This task is accomplished by exploiting the photon energy dependences of the photoemission cross-sections for various kinds of states. One useful dependence is that on atomic orbital [24]. Another method, called "resonance photoemission" spectroscopy (RESPES) [25], exploits cross-section resonances associated with certain core level absorption edges. As an example [23, 25] of RESPES relevant for this article, the 4f emission of rare earth atoms with unfilled 4f shells is greatly enhanced for photon energies near the 3d absorption edge by the photon absorption process near 880 eV, $3d^{10}4f^n \rightarrow 3d^94f^{n+1}$, followed coherently by the Auger emission process $3d^94f^{n+1} \rightarrow 3d^{10}4f^{n-1}\varepsilon_k$, where $\varepsilon_k$ is the outgoing photoelectron. RESPES at such high photon energy edges is also a way to overcome the problem of the generally reduced cross-section mentioned above. Two caveats for the use of RESPES are that the lineshape measured can by affected by contributions from incoherent Auger emission and that the Auger step of the coherent process may alter the relative intensities of different parts of the spectral function depending on the particular one-hole states of the spectral function and the particular virtual intermediate states of $3d^94f^{n+1}$ excited for the particular photon energy used. Nonetheless, used with care, RESPES has been enormously powerful in the study of strongly correlated electron systems. [25].

A recent advance in technique is that of making Fermi surface intensity maps [26], in which the electron detector is fixed at a kinetic energy corresponding to the Fermi energy for the photon energy being used, and the detector angles are then varied. The result can be interpreted as the intersection of the spherical **k**-space surface traversed with the Fermi surface. It is also possible to make such a map perpendicular to the sample surface by varying the photon energy and one detector angle.

2.3 ARPES instrumentation

The small energy dispersions and energy scales of many correlated electron materials impose stringent requirements on the **k** (angle) and energy resolutions, and on the sample temperature. Recent advances in instrumentation are promising. Laboratory gas discharge lamps, e.g. the Gammadata He lamp, and undulator-based synchrotron beamlines offer much higher photon intensities than before and thus enable photon energy resolutions ranging from < 2 meV for 20 eV photons to < 100 meV for 1000 eV photons. Particularly notable for high resolution at high photon energy is the twin-helical undulator beamline BL25SU of the SPring-8 synchrotron in Japan [23]. New electron detectors and energy analyzers offer an order of magnitude improvement in energy and angle resolutions, e.g. to ≈2 meV and ≈0.2°, respectively, combined with multi-channel energy and angle detection that greatly speed up data-taking and improve the signal to noise of the spectra. More elaborate sample cooling schemes are enabling sample temperatures ≈10K or less to be achieved.

## 3. Fermi liquid with translational symmetry

### 3.1 Luttinger's Fermi liquid formulation

Luttinger's treatment [27] of the Landau picture provides a description of general properties of the quasi-particle peak for a translationally symmetric metal. For a single non-degenerate band $G(\mathbf{k},\omega)$ is represented in terms of the many body self energy $\Sigma(\mathbf{k},\omega)$ as $[\omega - \varepsilon_\mathbf{k} - \Sigma(\mathbf{k},\omega)]^{-1}$ where $\varepsilon_\mathbf{k}$ is the non-interacting (or averaged interaction) single-particle energy and energies are referenced to the chemical potential $\mu$. For $\Sigma(\mathbf{k},\omega) = 0$ one has $\rho(\mathbf{k},\omega) = \delta(\omega - \varepsilon_\mathbf{k})$ so that the $\mathbf{k}$-summed spectral function is just the ordinary one electron density of states. For non-zero $\Sigma(\mathbf{k},\omega)$ one has $\rho(\mathbf{k},\omega) = (1/\pi)|\text{Im } \Sigma(\mathbf{k},\omega)/\{(\omega - \varepsilon_\mathbf{k} - \text{Re } \Sigma(\mathbf{k},\omega))\}^2 + \{\text{Im } \Sigma(\mathbf{k},\omega)\}^2\}|$, a Lorentzian lineshape if $\Sigma$ is $\omega$-independent, but in general more complex. Although it is a general concept, the self energy is most useful in systems where it can be found by perturbation theory. Luttinger justified the Landau picture under the presumption that for $\mathbf{k}$ near enough to the Fermi surface $\mathbf{k}_F$, the leading $\omega$-dependence of Im $\Sigma(\mathbf{k},\omega)$ is $-i\beta_\mathbf{k}\omega^2$, for which causality demands that Re $\Sigma(\mathbf{k},\omega)$ be $\alpha_\mathbf{k}\omega$. The spectrum $\rho(\mathbf{k},\omega)$ is then peaked around the shifted "quasi-particle" poles of $\omega = E_\mathbf{k}$ that solve $(\omega - \varepsilon_\mathbf{k} - \text{Re } \Sigma(\mathbf{k},\omega)) = 0$, but is broadened by Im $\Sigma$, which plays the role of an inverse lifetime. The truncated $\omega$-dependence of $\Sigma(\mathbf{k},\omega)$ renders the resulting $G(\mathbf{k},\omega)$ to be non-causal for large $\omega$ and so the resulting spectral function $\rho(\mathbf{k},\omega)$ is not valid for $\omega$ too far from $E_F$. Matho [28] has given a simple and useful expression for $\Sigma(\mathbf{k},\omega)$ that has the low energy Fermi liquid form but is causal and valid for all $\omega$. It can be remarked that a similar issue of causality arises for the Drude model of the optical conductivity of a metal. When the inverse lifetime is $\omega$-dependent, e.g. with the FL form, maintaining causality leads to a generalized Drude analysis [29] which can be viewed [30,31] as showing the need for a concomitant $\omega$-dependent, enhanced mass and which has been elegantly applied [32,33] in analyzing the optical spectra of various correlated electron systems.

The Fermi surface is defined by the $\mathbf{k}_F$-values for which $E_\mathbf{k} = \mu$ (taken to be zero here), and Im $\Sigma(\mathbf{k},\omega)$ goes to 0 for $\omega$ going to $\mu$ fast enough that the broadened peaks sharpen into delta functions with weights $Z_\mathbf{k} = (1 - \alpha_{\mathbf{k}F})^{-1}$, i.e. $\rho(\mathbf{k}_F,\omega=\mu) = Z_{\mathbf{k}F}\delta(\omega - \mu) +$ (a background function). It can be shown that $1 \geq |Z_\mathbf{k}| \geq 0$. The occupations $n_\mathbf{k}$ of states $\mathbf{k}$ are given by integrating $\rho(\mathbf{k},\omega)$ up to $\mu$ and are not simply 1 or 0 as they would be for a non-interacting system. For $\mathbf{k}$ inside the FS $n_\mathbf{k}$ can be less than 1 and for $\mathbf{k}$ outside the FS $n_\mathbf{k}$ need not be zero, reflecting the fact that the interaction admixes hole-particle pairs into the ground state. Nonetheless the Fermi surface remains well defined but with a discontinuity of $Z_\mathbf{k}$ rather than 1 as $\mathbf{k}$ crosses the Fermi surface and the $\delta$-function contribution to $n_\mathbf{k}$ is abruptly lost. The celebrated Luttinger counting theorem is that the volume contained inside the surface $E_\mathbf{k} = \mu$ is the same as that contained inside the surface $\varepsilon_\mathbf{k} = \mu$. In essence, the number of electrons is unchanged by the interaction and the counting of them remains the same so the FS volume is preserved. Luttinger also showed that the T-linear specific heat coefficient $\gamma$ is given by the free-electron formula but using the quasi-particle density of states of $E_\mathbf{k}$, i.e. $\gamma = (\pi^2/3)\, k_B^2\, [\Sigma_\mathbf{k}\, \delta(\mu - E_\mathbf{k})]$. For thermodynamics each eigenstate is given a weight of one and its spectral content as measured by the weight Z is unimportant. So relating $\gamma$ to the spectral function requires a factor $Z^1$, giving $\gamma = (\pi^2/3)\, k_B^2\, [\Sigma_\mathbf{k}\, Z_{\mathbf{k}F}^{-1}\, \rho(\mathbf{k}_F, \omega=\mu)]$. The last step appears to neglect the background function, but it can be shown [34] that the result is generally true. By defining an averaged $(Z^{-1})$ by $(Z^{-1})_{AV} = [\Sigma_\mathbf{k}\, Z_\mathbf{k}^{-1}\, \rho(\mathbf{k}_F, \omega=\mu)]/\rho_{LOC}(\omega=\mu)$ where the $\rho_{LOC}(\omega) = \Sigma_\mathbf{k}\, \rho(\mathbf{k},\omega)$ is the **k**-integrated, i.e. local, spectral weight, $\gamma$ can be given the useful form $\gamma = (\pi^2/3)\, k_B^2\, (Z^{-1})_{AV}\, \rho_{LOC}(\omega=\mu)$.

Finally, if $\Sigma(\mathbf{k},\omega) = \Sigma(\omega)$, i.e. is **k**-independent, then one can show further [35] that (a) the Fermi surface shape as well as its volume is unchanged by the interactions, (b) the local spectral function has the conservation property that $\rho_{LOC}(\omega=\mu) = \rho_{0,LOC}(\omega=\mu)$, so that (c) $\gamma = (\pi^2/3) k_B^2 [Z^{-1}]\rho_0(\omega=\mu) = \gamma_0/Z$, where the subscript '0' means 'unperturbed.' Thus the reduction of weight from 1 to Z is directly connected to the enhancement of the mass in the T-linear specific heat. One sees here the same generic causality-driven link between mass enhancement, involving the self-energy parameter $\alpha$, and a frequency-dependent lifetime, involving the self-energy parameter $\beta$, that occurs in the generalized Drude optical conductivity [29,30,31] mentioned above. These results are useful for making connection to the Anderson impurity model described below, because that model's spectral function satisfies the same Fermi energy sum rule as $\rho_{LOC}(\omega)$ and has the same general expression for $\gamma$ as that for a **k**-independent self energy. The common features suggest the possibility of representing a **k**-independent self energy using some suitably determined impurity Anderson model. That indeed turns out to be the case and is exploited in the theoretical technique known as dynamic mean field theory (DMFT), described in sub-section 5.2 below.

3.2 A Fermi liquid example: TiTe$_2$

TiTe$_2$ is a quasi-2d material whose transport properties show it to be a Fermi liquid [36]. Its electronic structure has a non-degenerate band crossing $E_F$ at an isolated place in **k**-space. This band is easily observed in ARPES and is widely studied as an example of Luttinger's FL spectral function. Fig. 1 shows the comparison of ARPES data and FL theory lineshapes from Ref. [37] in which the material was first proposed as a FL reference. The fitting parameter given in the figure is $\beta' = Z_F \beta$, where $Z_F$ is the quasi-particle weight on the FS, taken as **k**-independent in the data analysis. Further details of the experimental procedures and data analysis are set forth in Ref. [37]. Here we note only a few points. First, the lineshapes near $E_F$ are well described by the theory, but further from $E_F$ near the bottom of the dispersing band it is clear that the lineshape is more complex. This lineshape can actually be fit [38] by the Matho self energy if the self energy is allowed to be **k**-dependent. Second, the amplitudes of the various spectra have been normalized to be the same at the peak position. In fact the intensity falls sharply for **k** outside the FS, as would be expected, and the spectra for these k are fit by the part of the $k > k_F$ spectral function lying below $E_F$, i.e. the part mentioned above that gives rise to $n_\mathbf{k} > 0$ for $k > k_F$. Third, the least binding energy edge is dominated by the broadenings due to the experimental angle and energy resolutions and the temperature. Data utilizing the better resolutions recently available, at lower temperature, and also over a wider photon energy range to better determine the departure from

quasi-2-dimensionality, are being taken by various groups [39], but have not yet appeared in the literature.

4. The Anderson model

The Kondo behavior of the impurity [40] and lattice versions of the Anderson model underlies much of the thinking about heavy Fermion f-electron systems. This behavior provides a scenario for spin fluctuations as an emergent low energy scale property, controlled by high energy scale charge fluctuations, and generating the large heavy Fermion mass. The charge fluctuations are the model ingredients. The impurity model describes an $N_f$-fold degenerate impurity orbital with binding energy $\varepsilon_f$ and local Coulomb repulsion $U_{ff}$ hybridized by matrix element $V(\varepsilon)$ to non-interacting conduction band electrons with density of states $D(\varepsilon)$. $U_{ff}$ acts to energetically separate n-electron valence states $f^n$, $f^{n-1}$, etc., and $V(\varepsilon)$ enables valence fluctuations through electron transfer back and forth with the conduction band. The lattice model envisions a periodic array of such "impurities."

4.1 The Anderson impurity model and its quasi-particle peak

The impurity model applied to the f-electron of a nominally trivalent $f^1$ Ce ion [41] provides a useful example of its emergent Kondo properties, and of quasi-particle behavior as well. The former is viewed best by varying $V(\varepsilon)$ and the latter by varying $U_{ff}$. Suppose first that $V(\varepsilon)$ is zero and that $U_{ff}$ is non-zero and fixed. In the spectrum $\rho_f(\omega)$ to add and remove Ce f-electrons, below and above $\mu$ there are $4f^1 \rightarrow 4f^0$ ionization and $4f^1 \rightarrow 4f^2$ affinity peaks, roughly at $-|\varepsilon_f|$ and $-|\varepsilon_f|+ U_{ff}$, respectively, relative to $\mu$. Such a system has a local magnetic moment and a Curie law magnetic susceptibility $\chi(T)$. For $V(\varepsilon)$ not zero the low energy physics is profoundly different. The mixed f/conduction electron ground state is a singlet, and as temperature T decreases $\chi(T)$ changes from Curie-like to Pauli-like around a Kondo temperature $T_K$, the emergent low energy scale of the model. The T=0 f-spectrum, first calculated by Gunnarsson and Schönhammer [41], has now an additional peak, the "Kondo" or "Suhl-Abrikosov" resonance that occurs at $E_F$. Viewed this way the resonance is the spectral manifestation of the emergent Kondo properties and appears abruptly in the spectrum as V departs from zero, accompanying the discontinuous change in the character of the ground state, from magnetic moment presence to magnetic moment absence.

Varying $U_{ff}$ in the model reveals a specific example of the workings of quasi-particle physics at both high and low energies, with the resonance as the quasi-particle peak. Suppose initially that $U_{ff}$ is zero and that $V(\varepsilon)$ is non-zero but fixed. Electrons can be added and removed at the same energy, $\varepsilon_f$. The f-spectrum is then a delta function at $\varepsilon_f$, broadened by $V(\varepsilon)$ into an $N_f$-fold degenerate virtual bound state (VBS) cut by the chemical potential. The VBS broadening and the value of $\varepsilon_f$ relative to $\mu$ determines the total f-occupation $n_f$ (near 1 for $Ce^{3+}$). With equal occupation of all $N_f$ orbitals the magnetic moment is zero. This is the unperturbed reference state for this impurity Fermi liquid. $U_{ff}$ is now increased from zero and it is supposed that the magnetic moment nonetheless remains zero, i.e. that the ground state symmetry is unchanged. It has been shown [42] that the many-body form of the Friedel sum rule plays the same role in this impurity problem as does the Luttinger counting theorem for the case of translational symmetry. Specifically, if the ground state symmetry stays the same and if $\varepsilon_f$ is adjusted relative to $\mu$ to maintain the same $n_f$, then as $U_{ff}$ is increased from zero the magnitude of the spectral function at the chemical potential, $\rho_{ff}(\mu)$, is fixed to be the same $\rho_{0ff}(\mu)$ as it would be for a system with $U_{ff}=0$ and the same nonzero $V(\varepsilon)$. For large $U_{ff}$ the spectral function must display its atomic-like peaks at $-|\varepsilon_f|$ and $(-|\varepsilon_f|+U_{ff})$, and yet a quasi-particle peak of reduced width and weight, i.e. the resonance, must remain at $\mu$ to satisfy the Fermi energy sum rule. The weight $Z_{ff}$ of this coherent peak is related to an enhancement of the impurity contribution to the specific heat and so to an enhanced effective mass. Thus one can consider either the effect of nonzero V in causing a dramatic and abrupt low energy departure from the high energy atomic-like spectrum, or of nonzero $U_{ff}$ in narrowing the VBS and pushing spectral weight out to the atomic energies while maintaining at $E_F$ a quasi-particle peak. Yet another view is presented in sub-section 4.3 below.

Simple analytic results in the large degeneracy limit [41] for T=0 illustrate the basic features as follows. For $U_{ff}$ very large one considers $f^0/f^1$ mixed valence only, i.e. $1 \geq n_f$, and in the limit $N_f \to \infty$ and if $N_f\Delta(\varepsilon)=N_f(\pi\rho V^2)$ is constant, then $k_B T_K = E_F \exp(-1/J)$ where the Fermi energy is $E_F$ and the Kondo coupling constant is $J=N_f\Delta/\pi\varepsilon_f$. At T=0 the magnetic susceptibility per impurity is given by $\chi(0) = (n_f)C/T_K$, where $n_f$ is the f-state occupation and C is the Curie constant for the ionic ground state. With increasing T, as the magnetic moment is "unquenched" and its entropy evolves, there is a T-linear contribution $\gamma T$ to the specific heat, with $\gamma$ having the Fermi liquid form given above for a **k**-independent self energy, $(\pi^2/3)k_B^2 Z_{ff}^{-1}\rho_{ff}(\mu)$. The spectral function gives $\rho_{ff}(\mu) = \pi n_f^2/N_f\Delta$, which is indeed the value of $\rho_{0ff}(\mu)$ obtained for the $U_{ff}=0$ virtual bound state with occupation $n_f$, showing that the Friedel sum rule is satisfied and thus giving confidence in the calculation. $Z_{ff}$ is $(1-n_f)$ and one can show [41] that $(Z_{ff})^{-1} = 1/(1-n_f)= N_f\Delta/\pi n_f k_B T_K$, leading to $\gamma=\pi^2 k_B n_f/3T_K$ per impurity. If $T_K$ is small, i.e. the entropy evolves over a small temperature range, $\gamma$ can be very large, and so this is an important model for the large $\gamma$-values of heavy Fermion materials, which can be viewed as showing an enhanced conduction band mass m*.

4.2 The dense impurity ansatz for Ce compound 4f spectra: early work and $La_{1-x}Ce_xAl_2$

As described in various overview articles [43-46], work dating from 1981 [47] and continuing with increasing degrees of experimental and theoretical sophistication to the present has used the impurity model results to interpret local spectral and transport properties of concentrated cerium compounds, and to provide a basis for the Kondo volume collapse model [48,49] of the $\alpha$-$\gamma$ transition in Ce metal, described further in sub-section 4.4 below. The effort is sometimes referred to as the dense impurity ansatz. The dense impurity ansatz amounts to the assumption that the local spectral function for a lattice f-system is approximated by the impurity spectral function, i.e. $\rho_{LOC}(\omega) \approx \rho_{fIMP}(\omega)$ and that $(Z^{-1})_{AV} \approx (Z^{-1})_{IMP}$, where $\rho_{LOC}(\omega)$ and $(Z^{-1})_{AV}$ were defined above. Although there is no theoretical proof that this assumption is true, it is plausible that the impurity behavior could approximate the local aspects of a lattice system and the success of the scheme reported by various workers over the years is empiric evidence that this is so.

Fig. 2 shows an early comparison from Ref. [43] of impurity theory and low resolution experimental spectra from combined RESPES (photon energy at the Ce 4d edge) and X-ray IPES spectra of CeAl and CeNi$_2$, materials with very small and very large T$_K$'s, respectively. The details of the experiment and of the theoretical analysis are given in Ref. [43]. The experimental spectra show the dramatic development of the Fermi energy peak, first identified in 1983 [50] as the quasi-particle peak of the impurity model, and the theory curve shows the ability of the impurity theory to describe the essential features of the data. At the simplest level these features include the atomic-like ionization peaks 2 eV below E$_F$ and affinity peaks 5 eV above E$_F$, and the quasi-particle peak near E$_F$. The magnetic susceptibilities calculated from the impurity theory for the parameters extracted from the experimental spectra in Ref. [43] were found to be in reasonable agreement with the measured values. The detailed impurity model interpretation of Ce materials has continued vigorously to the present, with steady improvements in theory for T=0 [51-54] and T>0 [55], experiment [44,45, 56-58] and spectrum analysis [59,60]. Ref. [46] is a recent overview of the success and challenges of the approach, and of its relation to the lattice aspects, discussed next in connection with Fig. 3 and further in sub-sections 4.3 and 4.4 below.

Fig. 3 shows recent 20K PES data for polycrystalline samples of La$_{1-x}$Ce$_x$Al$_2$ with x=1 and x=0.04 which provide a direct experimental test of the dense impurity ansatz for a small T$_K$ system [61]. The conceptual strategy of the experiment exploits the Laves phase structure of CeAl$_2$, which has only 4 near Ce neighbors of a Ce site. For x=0.04 a fraction $(1-.04)^4 = 85\%$ of the Ce ions are completely isolated from other Ce ions if next-nearest neighbor Ce-Ce couplings can be neglected, so that the x=0.04 spectrum is dominated by the contributions of the isolated ions and can then be compared to the spectrum for x=1. The experimental strategy exploits the capability [23] of beamline 25SU at Japan's SPring8 synchrotron to perform photoemission with resolution of 100 meV at photon energies around the Ce 3d absorption edge near 880 eV. The large PES Ce 4f cross-section resonance at the Ce 3d edge is crucial for obtaining the x=0.04 spectrum and the high photon energy also minimizes the surface sensitivity of the measurement. For binding energy greater than 0.5 eV the spectrum is entirely x independent. The multiple $4f^1 \to 4f^0$ ionization peaks in this energy range reflect structure in the s-p-d density of conduction electron states to which the f$^0$ final states are hybridized. The peak 0.25 eV below E$_F$ is the "spin-orbit sideband" [55] of the quasi-particle peak and the peak nearest E$_F$ is a mix of the quasi-particle peak itself and weight from unresolved "crystal field sidebands" [55].

One can gain further insight into the quasi-particle weight at E$_F$ and the sidebands as follows. The quasi-particle one-hole final state has one conduction band hole and the original f-occupancy of the non-magnetic ground state, even though the peak occurs in the f-spectrum. Heuristically, one can visualize that the removal of the f-electron is followed by "hybridization screening," in which the initial f-hole is filled with a conduction band electron from E$_F$, thereby lowering the energy of the one-hole final state as much as possible and putting some f-spectral weight at E$_F$. For the atomic-like $4f^1 \to 4f^0$ ionization peak(s) below E$_F$ this screening process has not taken place even though the weight is spread into a VBS. The quasi-particle sidebands have final states in which hybridization screening has again restored the f-state occupancy to that of the ground state, but with a spin-orbit or crystal field excitation, i.e. a rearrangement of the internal degrees of freedom of the f-shell. For small T$_K$ the spectral weights of the spin-orbit and crystal field sidebands, first resolved by Y. Baer and collaborators [62,63], exceed that of the Kondo peak itself [41,43]. In the spectra of Fig. 3, the small reduction of the intensities of the two peaks relative to the $4f^1 \to 4f^0$ part, and the stronger decrease of the nearest E$_F$ weight relative to the spin-orbit sideband, are both expected within the impurity theory because of the known [64] reduction of $T_K$ from 4.8K for x=1 to 0.4K for x=0.04. This is a particularly nice test of the $T_K$ dependence of the spectrum since, unlike the situation in changing from compound to compound, the functional form of $V(\varepsilon)$ is essentially unchanged.

The $T_K$ values for this system are derived [64] from the T-linear specific heat coefficient, and the reduction is due to the decreased hybridization and consequent decreased coupling constant J that accompanies a volume expansion as the larger $La^{3+}$ ion replaces the smaller $Ce^{3+}$ ion. The reduction is opposite to expectations from the Nozieres' "exhaustion" idea [65], that in the simple Kondo lattice there are not enough conduction electrons to screen out the magnetic moments as occurs in the impurity model, essentially because only electrons located within the Kondo temperature of $E_F$ take part in the Kondo screening. The exhaustion problem is thus most serious in small $T_K$ systems and would imply a decreased $T_K$ in the lattice system relative to the impurity system. Numerical studies of the $N_f=2$ Anderson lattice [66] have given support to the exhaustion idea, but the behavior of $La_{1-x}Ce_xAl_2$ shows that other effects, perhaps the large degeneracy and the volume change, greatly outweigh any that are due to exhaustion.

4.3 Lattice aspects: f-electron FS exclusion above the Kondo temperature for $CeRu_2Si_2$

Relative to the impurity theory and the **k**-integrated photoemission that has been done to test it, both theory and experiment aimed at the **k**-dependence of the spectral function are just beginning. The spectral theory thus far [66] has been for models lacking the realism, e.g. the large degeneracy, which has been found to be important in analyzing the **k**-integrated spectra of Ce materials. Momentum-dependent effects have been reported in early ARPES studies of f-electron systems, e.g. high resolution He lamp excited valence band measurements [67,68}, amplitude variations of near $E_F$ 4f weight periodic with the Brillouin zone [69,70] and claims of "band-like" dispersion of f-states [71]. However, because of the technical barriers mentioned in section 2, ARPES has not provided enough FS detail for its potential to be realized. This section and the next describe recent advances [72] in ARPES studies of f-electron systems.

Application of the Luttinger theorem introduced in section 3.1 implies that, regardless of the route to a FL, if their magnetic moments are quenched then the f-electrons must be counted in the Fermi surface volume [73]. FS studies [74] using magneto-oscillatory techniques such as the de Haas van Alphen (dHvA) effect have verified that this is so. The most straightforward theory [75] for the FS utilizes the LDA band structure, including the f-electrons. A more sophisticated approach includes the f-electrons but renormalizes their scattering phase shifts at the FS to be Kondo-like. This procedure gives heavy masses and often makes only a minor perturbation of the original LDA FS [76]. A crucially important theoretical conjecture by Fulde and Zwicknagl [77,78] is that above $T_K$ where the f-electron moment is no longer quenched, the f-electrons should now be excluded from the FS volume. This idea is plausible but there is no proof having the rigor of the Luttinger theorem and in any case it would not hold experimentally if in fact Kondo physics were irrelevant in the heavy Fermion materials, as some workers propose. This section presents ARPES data verifying the Fulde/Zwicknagl conjecture.

The literature paradigm for low temperature f-electron inclusion in the FS volume comes from beautiful FS studies in the early 90's of the isostructural compounds $LaRu_2Si_2$, $CeRu_2Si_2$, and $CeRu_2Ge_2$ [79-81], using magneto-oscillatory (MO) techniques such as the de Haas-van Alphen (dHvA) effect. The cross-sections of panel (b) of Fig. 4 illustrate the LDA FS's of $f^0$ $LaRu_2Si_2$ and $f^1$ $CeRu_2Si_2$ [82]. Hole (electron) pockets are shown with bold (thin) lines. There are several pieces, including a large hole pocket centered around the Z-point of the Brillouin zone, shown in panel (a), smaller hole pockets centered around Z, a multiply connected electron sheet centered around Γ, and also a small electron pocket around Γ. The LDA bandstructure for $CeRu_2Si_2$ includes the 4f electron of the $Ce^{3+}$ ion in the FS. The electron is accommodated about 50% each in the multiply connected electron sheet, which expands and changes topology, and the large Z-point hole pocket, which shrinks to become the so called "pillow" of $CeRu_2Si_2$, shown in panel (a). These two pieces of Fermi surface have the large masses that account for the γ-value of this heavy-Fermion material. It is found [78] that dHvA data show all orbits expected within the renormalized LDA for $CeRu_2Si_2$, and show the expected size differences for the La and Ce systems. Further, $CeRu_2Si_2$ displays a metamagnetic transition under a sufficiently high magnetic field and $CeRu_2Ge_2$ is a ferromagnet. The dHvA experiments find that in these two situations where the f-moment is restored, the measured FS is like that of $LaRu_2Si_2$, again showing that the Luttinger theorem is obeyed [79,80].

MO techniques do not provide a global view of the energy and **k**-dependence of the electronic structure, and so are limited for distinguishing among different many-body scenarios of heavy mass formation. MO techniques also tend to be limited to low temperature and so have been silent about the crucial question of the FS volume above $T_K$. ARPES in principle provides this global view and is more easily done at high temperature than low temperature. The ARPES FS map of $LaRu_2Si_2$ in the upper part of panel (c) of Fig. 4 shows the large Z-point FS hole pocket in almost perfect agreement with the LDA prediction and with dHvA data. Smaller FS features around Γ and Z are visible but not resolved. Shown in the lower part of (c) is the ARPES map for $CeRu_2Si_2$ taken at 120K, well above its Kondo temperature of 20K. A careful comparison of the two maps [18] shows them to have nearly identical large Z-point hole pockets, as indicated by the arrows in the figure, and the so called "pillow," plus other predicted LDA features for $CeRu_2Si_2$, are not observed. Since the number of conduction electrons for the La and Ce compounds are the same if the Ce 4f electrons do not contribute to the FS volume, we conclude [18] that we have strong evidence of f-electron FS exclusion in $CeRu_2Si_2$ above its Kondo temperature. Note that this conclusion is a not a statement about the amount of 4f spectral weight observed in this part of the FS, although in fact ARPES measured by resonant photoemission shows it to be essentially unobservable, but rather on the counting of 4f electrons as implied by the size of this piece of FS.

Resonant ARPES reveals that parts of the FS show easily observable 4f weight at 120K. As presented in Ref. [18] this 4f weight is found in **k**-space in the vicinity of the low mass parts of the FS, the small hole pockets around Z and the small electron pocket around Γ. The angle integrated 4f spectrum at and above $T_K$ obtained directly [23] or by **k**-summing the resonant ARPES spectra [83] has the general appearance of the impurity-model spectra of Figs. 2 and 3. Although this finding may be plausible in view of the general theoretical correlation of smaller mass, larger energy scale and larger spectral weight, there is presently no spectral theory of the Anderson lattice capable of providing insight into ARPES data at this level of detail. The so called "LDA + DMFT" theory discussed in section 5.2 below is promising for further analysis of such data for systems where the Fermi surface is well described by the LDA.

Another important point concerns the bulk sensitivity of the CeRu$_2$Si$_2$ ARPES data of Ref. [18]. These spectra were not measured at the high photon energies that enhance bulk sensitivity and which are generally necessary [23] for bulk sensitive Ce 4f studies. However it was found [18, 78] that some cleaved surfaces yield 4f spectra at low photon energies, e.g. in RESPES at the Ce 4d edge, that are essentially the same as those at high photon energy, e.g. RESPES at the Ce 3d edge. Such surfaces were used to obtain the data of Ref. [18]. The explanation [83] probably lies in the particular crystal structure of CeRu$_2$Si$_2$, which admits of two cleavages, one that exposes Ce at the top layer and another that places Ce in a buried layer. The latter situation is expected to enable bulk sensitive studies even at lower photon energies.

4.4  Lattice aspects: Refined view of the Kondo volume collapse model

The confirmation of the Fulde/Zwicknagl conjecture described in the preceding sub-section affords a refined view of the Kondo volume collapse model [48,49] of the paradigmatic $\alpha$-$\gamma$ transition in Ce metal. In this model the smaller volume of the low temperature collapsed $\alpha$-phase increases its Kondo coupling constant so that its Kondo temperature $T_K(\alpha)$ is much larger than that $T_K(\gamma)$ of the expanded high temperature $\gamma$-phase. The change [60] of the angle-integrated 4f spectrum across the transition is then a less dramatic version of the difference between the two spectra shown in Fig. 2. The Ce magnetic moments are quenched in the $\alpha$-phase because $T < T_K(\alpha)$ and are restored in the $\gamma$-phase because $T > T_K(\gamma)$. The low temperature phase is stabilized by the large Kondo binding energy which more than offsets the increase of the non-4f contribution to the elastic energy, the latter being caused by the volume compression. The high temperature phase is stabilized by the entropy of the Ce moments and by a decrease of the non-4f elastic energy as the lattice expands, these two together offsetting the decrease of the Kondo binding energy in the free energy balance at the transition. The coming and going of the Ce moments in the two phases implies that the Ce f-electrons are included in the Fermi surface volume in the $\alpha$ phase and excluded in the $\gamma$-phase. But if the $\gamma$-phase could be cooled to well below its Kondo temperature the Fermi surface volume would then increase again to contain the 4f electron. Thus there is no fundamental difference between the two phases beyond the difference in their Kondo temperatures. The inclusion or not of an electron in the Fermi surface volume is perhaps the only way in which precision can be brought to the vague but commonly used terms, "itinerant" and "localized," respectively.

4.5  Lattice aspects: ARPES spectra of URu$_2$Si$_2$

Theoretical treatments of a simplified Anderson lattice Hamiltonian [66, 84, 85] suggest a model with the f-level renormalized from its atomic energy $\varepsilon_f$ well below $E_F$ to an energy $\varepsilon_f'$ just above $E_F$, rather like the Kondo resonance of the impurity model, and having renormalized hybridization to the s-p-d band structure. The renormalization of the hybridization is also linked to the Kondo effect. The simplified model has only a single "f" and a single "d" conduction band. Two band mixing very near $E_F$ shifts the $k_F$ value of the conduction band to a new value $k_F'$ and enhances the effective mass at $E_F$ but the general FS topology appears to be largely determined by the underlying s-p-d band structure that would obtain in the absence of hybridization to the f-states. Panel (d) of Fig. 5 shows the two band mixing model and panel (f) shows the distribution of "f" and "d" weight that is implied. A striking general prediction of this picture is the confinement of nearly dispersionless f-spectral weight inside a d-electron hole pocket.

Panel (a) shows resonant ARPES spectra from Ref. [18] for the heavy Fermion material URu$_2$Si$_2$ [31, 86], obtained along the $\Gamma$-X line at a photon energy near the U 4d edge where U 5f weight is resonantly enhanced, and panel (b) shows an intensity plot of the data. Panel (c) shows a similar intensity plot of ARPES data along this line but for a photon energy where the 5f weight is suppressed so that presence of a d-character hole pocket centered on X is easily seen. Panels (b) and (c) show the confinement of f-weight inside the d-electron hole pocket, with striking similarity to the model weight distribution of panel (e). The ARPES spectra of CeRu$_2$Si$_2$ discussed above display very similar f confinement to the hole pockets around the Z-point.

An aspect of the resonant ARPES spectra obtained for URu$_2$Si$_2$ but not presented here is that the LDA bandstructure shows poor agreement with the ARPES spectra in the vicinity of $E_F$, an example being that the d-band hole pocket at X is not present in the LDA band structure. A general conclusion of the discussion in Ref [18] is that the LDA bandstructure of uranium materials tends to accommodate the two or three 5f electrons by the presence of 5f bands that cross $E_F$ and change the FS topology considerably from that of an $f^0$ analog material like ThRu$_2$Si$_2$. Thus for uranium materials the functioning of the LDA appears dissimilar to that of the simple Anderson lattice picture of Fig. 5. In contrast, apart from the missing low energy scale, the general functioning of the LDA for CeRu$_2$Si$_2$ is similar to that of the simple Anderson lattice picture in that the LDA 4f bands lie essentially above $E_F$ and hybridize to the s-p-d-band structure with concomitant shifts of the FS crossings that lead to the changes of FS sizes whereby the one 4f electron per Ce is accommodated.

Another important finding [18] in the resonant ARPES spectra obtained for URu$_2$Si$_2$ is that the X-point 5f spectrum of panel (b) is very T-dependent, in contrast to the T independence of the angle integrated spectrum, which is dominated by T-independent contributions in most of the Brillouin zone. The onset temperature of the observed change correlates well with the onset temperature of changes observed in transport and in the optical conductivity [31], often interpreted as a Kondo temperature.

<u>5.0 The Hubbard model</u>

The simplest one-band Hubbard model [87] supposes a lattice with s-like atomic or Wannier orbitals on each site. A tight-binding band is generated by hopping matrix elements 't' between nearest neighbors and electron correlations are included by a Coulomb repulsion 'U' at each site. The model offers many possibilities for generalization, including more orbitals per site, more distant neighbor hopping, differing Coulomb interactions for each orbital, Coulomb repulsion between electrons on different sites, etc.

5.1 Mott-Hubbard transition

The simplest one-band model with a filling of one electron per site suffices to pose the question of the Mott-Hubbard metal to insulator transition (MIT) [88] with varying t/U. For U=0 and non-zero t the system has a half-filled band and is a paramagnetic metal (PM). For t=0 and non-zero U the atomic valence states $s^0$, $s^1$ and $s^2$ are energetically separated and so the spectrum to add and remove electrons is, much as described above for the interacting electrons of the Anderson model with V=0, split by U into two atomic-like peaks $s^1 \rightarrow s^0$ and $s^1 \rightarrow s^2$ on either side of the chemical potential. But lacking the non-interacting electron gas of the Anderson lattice model, this Hubbard system is then an insulator with local moments associated with the half-filled local orbitals. Non-zero but small t induces antiferromagnetic Anderson superexchange coupling [89] between the moments, resulting in an antiferromagnetic insulator (AFI) below a high temperature paramagnetic insulator (PI) phase. Increasing t/U must eventually reach the PM phase. There has been much effort over the years to identify experimental Mott-transition systems and to develop a detailed theory of the transition. The simplest spectroscopic picture for the transition is that the sharp peaks for large U broaden with increasing t until the gap is filled in. But new theory described next suggests quite a different picture.

5.2 Dynamic mean field theory

The dynamic mean field theory (DMFT) [90,91] is a relatively recent development in many-body theory that has provided, among other things, a new view of the Mott transition. The theory can be viewed either as the exact result of considering a system in the limit of infinite dimension, or of assuming that the **k**-dependence of the single-particle self energy of a non-infinite system can be neglected. As remarked above, the latter aspect makes it plausible that the theory can be formally mapped to a effective Anderson impurity problem and this is so, with the quite non-trivial addition that there is a self consistency condition between the "impurity" and the "electron gas" to which it is hybridized. In this approach the quasi-particle peak of the single-particle spectrum is described by the "Kondo resonance" of the impurity model. It is not that the model literally displays Kondo physics, but that the Kondo resonance has the right formal properties to describe general quasi-particle physics for a system with a **k**-independent self energy. Nonetheless the general Luttinger sum rule ideas [73] imply an underlying connection to magnetic moment suppression.

In application to the Mott transition [91] the self-consistency condition allows the system to bootstrap a narrow quasi-particle peak of reduced weight Z into the gap induced by the Coulomb interaction. I.e., the quasi-particle peaks coming from every site of the lattice establish together a state density at $E_F$ that is then used by any particular site to form the quasi-particle peak via the Kondo-like mechanism of the impurity model. The weight of the peak is taken from the weights of the states on either side of the correlation gap. The chemical potential lies in the quasi-particle peak. Essentially, the weight Z decreases with decreasing t/U and the Mott transition occurs when the weight Z goes to zero at some critical value of t/U. Given the central role of the impurity theory in this description it is no surprise that there is a similarity to the change of quasi-particle weight in the Kondo volume collapse model [48,49] of the cerium α-γ transition described in sub-section 4.4 above.

Two further theoretical developments are very promising. One is the generalization of the DMFT impurity problem to a local cluster [92,93], i.e. the "cluster DMFT," which gives a means for computing **k**-dependent corrections to the DMFT self energy. Another is the mating of the DMFT and the LDA [94]. This mating offers the possibility of comparing experimental results to theoretical spectral functions that include both a realistic description of the underlying electronic structure of a solid and many-body corrections that are exact to the extent that quasi-particle physics with a **k**-independent self energy obtains. Extensions of the "LDA+DMFT" to "LDA+cluster DMFT" are in progress but no results have yet been published..

5.3 PES view of the MIT in $(V_{1-x}Cr_x)_2O_3$

$(V_{1-x}Cr_x)_2O_3$ is the most well known and well studied MIT system [88]. In contrast to the well documented charge transfer insulator character of NiO [95] and of the superconducting cuprates, $V_2O_3$ appears to lie just to the charge transfer side of the boundary between charge transfer and Mott-Hubbard insulators in the Zaanen-Sawatzky-Allen classification scheme [96] for transition metal compounds. As a function of temperature and either x or pressure, PM, PI and AFI phases are realized. Because it does not entail magnetic ordering, the PM to PI transition accessed with changing x and/or pressure is taken as a paradigm example of a Mott-Hubbard MIT. There have been various past photoemission studies of the $(V_{1-x}Cr_x)_2O_3$ phase transitions [97-100]. For the few sets of data where the energy resolution has been adequate, e.g. [100] 100 meV or less, the photon energy has been relatively low, e.g. 110 eV or less, and in consequence the degree of surface sensitivity has been in the range for which surface effects have been found [101] in the PES spectra of other oxide MIT systems. Recent work [102] at beamline BL25SU of the SPring8 synchrotron in Japan has yielded the first well resolved high photon energy (and therefore very bulk sensitive) angle integrated photoemission spectra for all phases of $(V_{1-x}Cr_x)_2O_3$. Fig. 6 shows data for a composition, x=1.2%, in which all three phases can be accessed by varying the temperature, as marked in the inset. The overall energy resolution is about 90 meV and the photon energy is 500 eV, slightly below the V 2p→3d absorption edge. Although a large resonant cross-section enhancement of the V 3d spectrum occurs for photon energies just above the V 2p edge, the presence of accompanying incoherent Auger emission especially in the PM phase makes the off-resonance spectrum a better representation of the V 3d spectral function. Comparison to on-resonance spectra (not shown here) shows however that in the binding energy range roughly 1.5 eV to 2.5 eV of the 500 eV spectra there is a very small contribution from Cr 3d states. The Fermi edge of the PM phase and the somewhat differing gaps of the PI and AFI phases are easily seen in the spectra. An interesting aspect of the data is that the PI and AFI phase gaps, which are greater than or equal to their values on the PES side of $E_F$, are much larger than the energy scale, roughly 25 meV, that one would infer from the MIT transition temperatures.

The phase diagram of $(V_{1-x}Cr_x)_2O_3$ has traditionally been interpreted [103] within the framework of the one band Hubbard model. In the DMFT view of the transition the spectral weight that appears at $E_F$ in the PM phase has been transferred from the atomic-like insulator peaks to the bootstrapped quasi-particle peak. The DMFT view provides a possible rationalization of the large energy gap, namely that the energy scale of the MIT transition temperature is not that of the gap itself, which entails U, but instead involves the smaller energy scale of the quasi-particle peak, analogous the workings of the Kondo volume collapse model for Ce where the Kondo temperature, rather than the charge fluctuation energies, sets the energy scale for the transition. In assessing the DMFT theory for $V_2O_3$ it is important to take cognizance of recent work [104,105] showing that a more realistic many-band description is essential for the physics of the material. The first many-body spectral theory including such realism has recently been given using the LDA+DMFT method [106]. A comparison of the spectral functions obtained thereby to the data obtained in the SPring8 experiments is given elsewhere [102]. Another new line of theoretical development is an exploration of the possible importance of c-axis pairs of vanadium ions as the basic underlying unit of the electronic structure of the insulating phases, using both the model Hamiltonian [107,108] and the LDA+cluster DMFT [109] approaches. Such a proposal, based on optical and other properties of $(V_{1-x}Cr_x)_2O_3$ and $(V_{1-x}Al_x)_2O_3$, was put forth some time ago [110].

6.0 Non-Fermi liquid systems

At least two categories of active NFL study can be identified. One is the studies [111-113] of rare earth and actinide and transition metal alloys and compounds that display signatures of NFL behaviors initially identified as transport properties or neutron scattering spectra which are anomalous relative to the FL behavior of the usual one-channel Kondo effect. Some of these materials are alloys with the possibility of disorder and impurity effects, but others are pure materials with translational symmetry. Mechanisms for the NFL behavior that have been considered include the multichannel Kondo effect [15,114], various effects of disorder [115], and the quantum critical behavior found in proximity to a quantum phase transition [116]. Although the "Fermi level tuning" discovered by photoemission played a very important enabling role [117] in motivating transport studies [118,119] of $Y_{1-x}U_xPd_3$, the first such material discovered, thus far spectra that are directly relevant to the NFL properties of these materials have not been obtained and so they are not discussed here. Photoemission has, however, been directly relevant for another important category, studies of quasi-low dimensional materials, discussed next.

6.1 Electron fractionalization and the Luttinger Liquid

The behavior of a 1-d interacting electron gas provides a paradigm for a dramatic departure from quasi-particle physics, called electron fractionalization [120,121]. This behavior can be analytically demonstrated for the 1-d Tomonaga-Luttinger (TL) model [122,123], which features a Hamiltonian having a truncated Coulomb interaction such that the Hamiltonian can be exactly diagonalized by transforming to spin and charge density operators. One sees thereby that the only excitations of the system consist of independently dispersing spin and charge density waves called spinons and holons, having velocities $v_s$ and $v_c$, respectively, that differ both from one another and from the underlying single-particle Fermi velocity $v_F$. Unlike the workings of the FL model, where electron-electron interactions make effective masses larger, $v_s$ and $v_c$ can be larger than $v_F$. There are no quasi-particle excitations and the addition or removal of an electron results entirely in the production of multiple spinons and holons whose energies and momenta must satisfy 1-d kinematics to produce the energy and momentum of the added or removed electron. The electron is then said to be fractionalized. The TL model is thought to give the generic low energy behavior for all 1-d models of interacting electrons, including ones that cannot be solved analytically, and the behavior has been given the name [13] "Luttinger liquid" (LL). The TL model and its LL behavior are characterized by $v_s, v_c, v_F$, and also a fourth parameter, the anomalous dimension $\alpha$, discussed below. The low energy mapping of a particular model onto the TL model, and the relations among the four parameters, vary with the details of the interaction term of the model under consideration.

In consequence of fractionalization, the predicted ARPES lineshape [124-126] is much different from the broadened quasi-particle peak of a FL. Instead there are two dispersing features associated with the two kinds of component density waves. These features display singular behavior with power law tails characterized by the anomalous dimension $\alpha$. The lowest energy feature of the spectrum, typically the spinon-related peak that disperses with velocity $v_s$, is either an edge or peak singularity depending on whether $\alpha$ is greater than or less than ½ respectively. The most counterintuitive aspect of the lineshape is that the **k**-summed spectral function $\rho_{LOC}(\omega)$ decreases to zero as a power law in $\alpha$ as the magnitude of the binding energy approaches the chemical potential even though the system is metallic. From the discussion of section 3.1 above, it is immediately clear that the self energy must be highly **k**-dependent since otherwise $\rho_{LOC}(\mu)$ would have the non-zero value of the non-interacting system. In fact the self energy could not be obtained by perturbation theory on the non-interacting system and so is not a very useful concept for this situation.. General power law behavior also appears in other correlation functions for the system and can be seen as a manifestation of quantum criticality [116] in which the only energy scale is that of temperature. Orgad [126] has exploited this viewpoint recently in obtaining the temperature dependence of the single-particle spectral function of the TL model.

Interest in the LL was greatly intensified by Anderson's proposal [127] that the ARPES lineshapes and certain other properties of the superconducting cuprates signaled in a general way the possibility of LL behavior in quasi 2-d systems, and by pioneering PES studies [128-130] of quasi-1-d metals that found power law onsets to $E_F$ rather than a Fermi edge. A complication for interpreting the PES data is that below a transition temperature $T_{CDW}$ the quasi-1-d metals studied display static charge density wave (CDW) formation which gaps the quasi-1-d FS and produces an insulator. Especially in 1-d, strong CDW fluctuations involving electron-phonon interactions above $T_{CDW}$ can cause the PES lineshape to have NFL behavior that mimics that of the LL, in particular that there can be a pseudogap [131-133] such that the weight at $E_F$ is greatly suppressed. This situation has motivated ARPES activity [17] on low-dimensional non-cuprate materials with the intent of elucidating CDW behavior, of searching for a paradigm having the distinctive features of the LL lineshape, and of seeing connections to cuprate spectra. The finding of the FL paradigm $TiTe_2$ discussed in section 3.2 above was part of this general program, which has now progressed also to success with the other two goals, as described next.

6.2 LL paradigm $Li_{0.9}Mo_6O_{17}$

To date the "Li purple bronze," $Li_{0.9}Mo_6O_{17}$ is the only quasi-1-d metal known to display ARPES spectra [17,134,135] that compare favorably to the LL lineshape. $Li_{0.9}Mo_6O_{17}$ has two low-T- phase transitions, one at 1.9K to superconductivity and another at $T_X$=24K that is not understood. The transition produces a small hump in the specific heat [136] and a resistivity uprise with lowering T [137]. However there is no hint of a gap in the optical reflectivity [138] down to 1 meV, or in the T dependence of the magnetic susceptibility [139], and repeated X-ray diffraction studies [140] have found no evidence for a CDW or a spin density wave (SDW). If the resistivity uprise is interpreted as the opening of a gap $\Delta$, a value of $2\Delta = 0.3$ meV [137] is obtained, too small to relate to $T_X$ using CDW theory. An alternate explanation [139] for the uprise is localization effects associated with quasi-1-d disorder.

PES spectra (not shown here) [17] for $Li_{0.9}Mo_6O_{17}$ above $T_X$ display a power law onset to $E_F$ rather than a Fermi edge. Fitting to the T-dependent LL lineshape theory [126] developed only recently yields $\alpha$=0.9 and for T decreasing below $T_X$ there is only a slight sharpening of the edge but neither a gap nor a Fermi edge is observed. As shown schematically in panel (a) of Fig. 7 and found experimentally [17], the Li purple bronze has a total of four bands within 1 eV of $E_F$, two of which (C,D) converge to cross $E_F$ at the same **k**-value in the quasi-1-d ($\Gamma$-Y) direction. In agreement with band theory [141], a FS map (not shown here) [134] reveals that the electronic structure near $E_F$ is highly 1-d. Panel (b) shows ARPES spectra along the $\Gamma$-Y direction. These data show all four bands, but for the particular experimental conditions the intensity of one of the bands (D) crossing $E_F$ is greatly suppressed. Panel (c) shows the data of (b) overplotted to emphasize the differing dispersions of the leading edge and the peak of band C, attributed to the spinon and holon components of the TL lineshape, respectively. The theoretical TL lineshapes of panel (d) are calculated for the temperature T=250K of the data. The $\alpha$ value of 0.9 is chosen, as in previous work, to simulate the amount of $E_F$ weight at the crossing relative to the maximum peak height, and agrees nicely with the value inferred from the PES $E_F$ onset. The $v_c/v_s$ value of 2 is used as in Ref. [135], where it was already noted that the improved angle resolution relative to that of the earliest work gives better resolution of the spinon edge and leads to a $v_c/v_s$ value changed from the early [134] value of 5. Overall the agreement between the data and the theory is excellent. To show that two dispersing features with difference velocities have indeed been observed, Fig. 8 compares the Li purple bronze data in panel (a) with TL model lineshapes for varying $v_c/v_s$ in panels (b) through (f). The underlying Fermi velocity is chosen so that the holon peak dispersion is always the same and is matched to the experimental peak movement. It is apparent by eye that the spinon edge movement is too fast in panels (b) and (c) and too slow in panels (e) and (f) to match that of the experimental data.

Fig. 9 illustrates the remarkable difference in the analytic properties of the FL and TL lineshapes found for $TiTe_2$ and $Li_{0.9}Mo_6O_{17}$, respectively, as shown by **k**-integration of the near $E_F$ ARPES lineshape data over a progressively larger **k**-range relative to $k_F$. Although the k=$k_F$ spectra for both have the general appearance of an edge, the former **k**-integrates to a Fermi edge whereas the latter **k**-integrates to the power law observed in the PES spectrum taken in an angle integrated mode.

6.3 Generalized Signatures of Electron Fractionalization

The results for $Li_{0.9}Mo_6O_{17}$ are part of a larger study of several low-d molybdenum bronzes, including also the "K blue bronze," quasi-1-d $K_{0.3}MoO_3$, and the "Na purple bronze," quasi-2-d $NaMo_6O_{17}$.

Both materials show [142] CDW formation below transition temperatures of 180K and 80K, respectively, and CDW fluctuation effects have been identified in X-ray diffraction above their CDW transition temperatures up to room temperature [143]. Above its CDW transition temperature the dc magnetic susceptibility of the blue bronze is activated [144], in contrast to its metallic dc electrical conductivity. This contrast suggests [145] the spin-charge fractionalization of the quasi-1-d Luther-Emery (LE) model [146], in which an electron-electron interaction channel additional to that of the TL model results in a gap in the spin excitations but not the charge excitations. The Na purple bronze is notable for its "hidden 1-d" FS [147,148] arising from planes that contain three weakly coupled chains mutually oriented at 120 degrees. It is thus an interesting and valuable bridge material between one and two dimensions.

Both the blue and the Na purple bronzes yield sharp FS maps in good general agreement with expectations from band theory [134,149]. However the ARPES lineshapes of the dispersing excitations that define their FS's are definitely of a NFL character. Their NFL lineshapes cannot be described by either the TL model or by CDW fluctuation models, and at the moment a detailed theoretical lineshape for the LE model is not available. Nonetheless their lineshapes share with those of $Li_{0.9}Mo_6O_{17}$ and also the quasi 2-d superconducting cuprates four commonalities that are generic to models that display fractionalization, e.g. the TL and LE models, and to related ideas. These commonalities are termed 'generalized signatures of electron fractionalization' and are presented in a recent paper [150].

One signature that is illustrated in Fig. 9 is the lack of a Fermi edge in angle-summed spectra and an associated anomalous dimension $\alpha$. Non-zero $\alpha$ is observed also for the K purple and the blue bronze. A second signature, as illustrated in Fig. 7 for the Li purple bronze, is the presence of two or more objects in the dispersing ARPES lineshapes. This signature has not been identified for the blue bronze, but here the situation is complicated by the presence of the spin gap and the current lack of LE model spectra. For the Na purple bronze one observes two independent components, both the dispersing peaks that define the FS [149, 150] and an equally large amount of **k**-independent weight that is neatly confined to the bandwidth of the dispersing peaks and hence cannot have an extrinsic origin such as the inelastic scattering of photoelectrons that causes the usual PES inelastic "background." We propose [151] that the **k**-independent weight arises from heavy scattering of holons by the well known charge disorder of the Na ions whereas the dispersing peaks are spinons, unscattered because they do not see the charge disorder. There is a striking similarity between these aspects of the ARPES spectra of the Na purple bronze [149,150] and those of the superconducting cuprates, e.g. $Bi_2Sr_2CaCu_2O_{8+\delta}$ [152]. A third signature [153] is that even though the FS is sharp and well defined, nonetheless the ARPES spectra for the Fermi surface **k**-values are very broad in energy due to their fractionalized component character. This signature is observed in Fig. 7 for the Li purple bronze, is also present in the blue bronze spectra, and is seen to be present for the Na purple bronze [149,150] and [153] for $Bi_2Sr_2CaCu_2O_{8+\delta}$ once it is appreciated [154,155] that the **k**-independent weights of their ARPES spectra are intrinsic. The last signature is a lineshape visualization devised by Anderson and Ren [156] for the tails of cuprate ARPES spectra and associated with the anomalous dimension $\alpha$. The Anderson-Ren lineshape is also found for the ARPES spectral tails of all three bronzes. These general results give strong support to the hypothesis that, even though present models are not adequate to describe most quasi-low-d materials, fractionalization is nonetheless the correct underlying NFL physics.

7.0 Outlook

The spectra presented in this paper are but a hint of what lies ahead for photoemission studies of electrons in condensed matter. Much can be done with the present capabilities of the technique and the prospects are very good for the further technical advances that are needed. The varied scenarios whereby the beautiful Fermi liquid paradigm is realized even in strongly correlated systems are being elucidated with unprecedented clarity and detail, and the appreciation for the Fermi liquid gained thereby only serves to heighten the great intrigue of exploring interesting ways in which the paradigm can fail.

Acknowledgments

It is a pleasure to acknowledge many years of stimulating and rewarding collaborations with numerous colleagues, many of which appear in the reference list for this paper. I am grateful to the Advanced Light Source of the Lawrence Berkeley National Laboratory for gracious hospitality over the period of a one-year (2002) sabbatical visit, during which time this paper was written. Work at the University of Michigan was supported by the U.S. DoE under Contract No. DE-FG02-90ER45416 and by the U.S. NSF under Grant No. DMR-99-71611. Work at SPring8 was supported by a Grant-in-Aid for COE Research (10CE2004) of MEXT, Japan and JASRI (No. 2000B0335-NS-np). The SRC is supported by the U.S. NSF Grant No. DMR 95-31009 and the ALS by the U.S. DoE under Contract No. DE-AC03-76SF00098.

Figure Captions

1. Fermi liquid line shape fits to ARPES spectra of an isolated band crossing $E_F$ in TiTe$_2$ [from Ref. 37].

2. Early comparison of Anderson impurity model spectral theory to Ce 4f RESPES/BIS spectra of small and large $T_K$ materials, showing ionization and affinity peaks well above and below $E_F$ and growth from small $T_K$ to large $T_K$ of quasi-particle peak (Kondo/Suhl-Abrikosov resonance) just above $E_F$. [from Ref. 43]

3. Nearly identical Ce 4f spectra of a dilute and concentrated Ce system, obtained by RESPES at the Ce 3d edge with the Kondo resonance spin-orbit sideband at 0.25 eV resolved, giving direct experimental verification of the "dense impurity ansatz" for a small $T_K$ cerium material. The slight decrease of the sideband and the larger decrease of the $E_F$ intensity with dilution are due to a known decrease of $T_K$ arising from volume expansion. [from Ref. 61].

4. ARPES Fermi surface map for T = 120K $\gg$ $T_K$ = 20K of CeRu$_2$Si$_2$ compared to ARPES map of LaRu$_2$Si$_2$, showing same size of large hole surface contour for both compounds. Absence of reduced hole surface size due to Ce 4f electron inclusion, as predicted by LDA and observed at very low T in dHvA experiments, is evidence of Ce 4f electron exclusion from Fermi surface at T $\gg$ $T_K$, as proposed in Refs. [77,78]. [from Ref. 18].

5. Summary of below (102 eV) and above (108 eV and 112 eV) resonance ARPES data in panels (a)-(c) for heavy Fermion material URu$_2$Si$_2$, showing confinement of 5f weight (panel (b)) to interior of Ru d-band hole pocket (panel (c)), as in simple 2-band ansatz for the Anderson lattice model where a Kondo-renormalized f-state at $\varepsilon_f'$ hybridizes to the d-band, as shown in panels (d) and (e). [from Ref. 18]

6. Bulk sensitive PES spectra taken at photon energy 500 eV, showing V 3d valence band in all three phases of MI transition material (V$_{0.982}$Cr$_{0.018}$)$_2$O$_3$. Paramagnetic metal (PM) phase weight near $E_F$ could be quasi-particle weight transferred from Hubbard peaks of the antiferromagnetic insulating (AFI) or paramagnetic insulating (PI) phases, as in the dynamic mean field theory [91, 106] of the MI transition. [unpublished data from collaboration of Ref. 102].

7. (a) schematic band structure of quasi-1-d metal Li$_{0.9}$Mo$_6$O$_{17}$; (b) ARPES spectra showing all four bands A-D; (c) overlaid plot of ARPES spectra for comparison of band C lineshape to spectral theory [Ref. 126] of Tomonaga-Luttinger model in panel (d), showing lack of Fermi liquid quasi-particle due to Luttinger-liquid-like electron fractionalization, presently the only such ARPES example. [from Refs. 135 and 150].

8. TL model spectra from theory of Ref. [126] in panels (b) to (f) compared to Li$_{0.9}$Mo$_6$O$_{17}$ ARPES data of panel (a) and Fig. 7 (c) to show sensitivity of Tomonaga-Luttinger description of data to choice of ratio of velocities of holon peaks and spinon edges. Holon peak dispersion is held constant and matched to experimental peak dispersion for ease of comparison of spinon edge dispersions. The lineshapes of panel (d) and Fig. 7 (d) provide an excellent description of the data.

9. **k** integration of ARPES spectra of dispersing excitation defining the Fermi surface, over a progressively larger range of **k** away from **k**$_F$, evolves to a Fermi edge in TiTe$_2$ and a power law onset in Li$_{0.9}$Mo$_6$O$_{17}$, even though the **k**$_F$ spectra of both are edge-like at $E_F$, thus showing for the latter a counter-intuitive signature of quasi-particle absence due to electron fractionalization.

**Fig. 1**

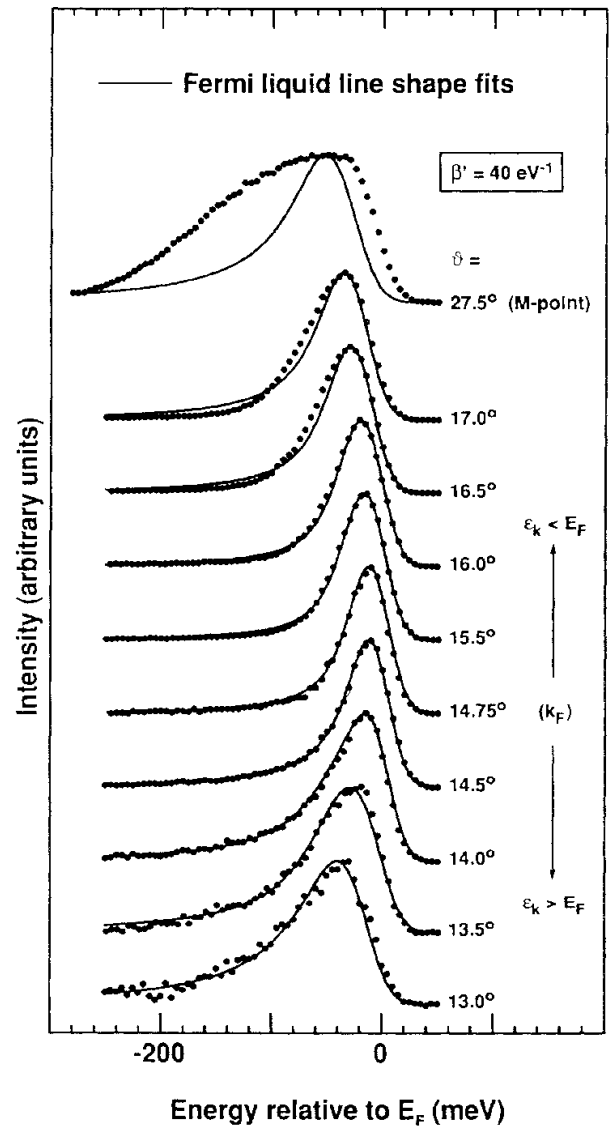

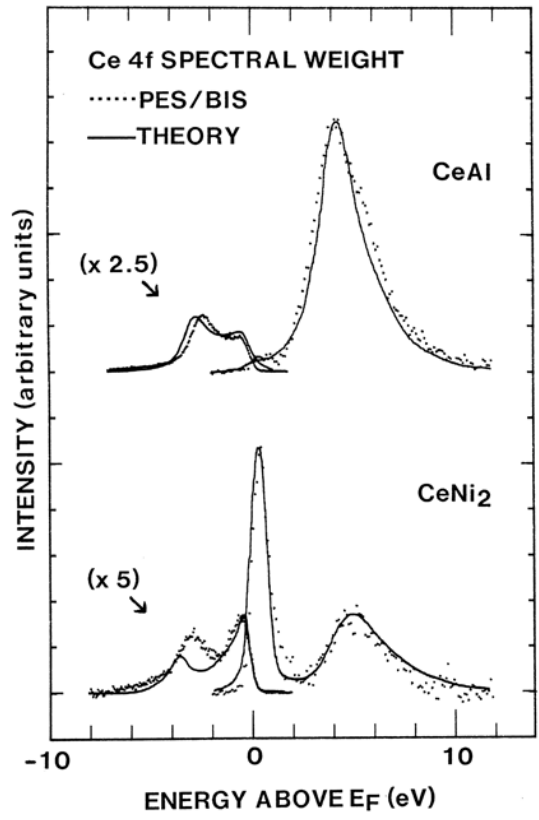

Fig. 2

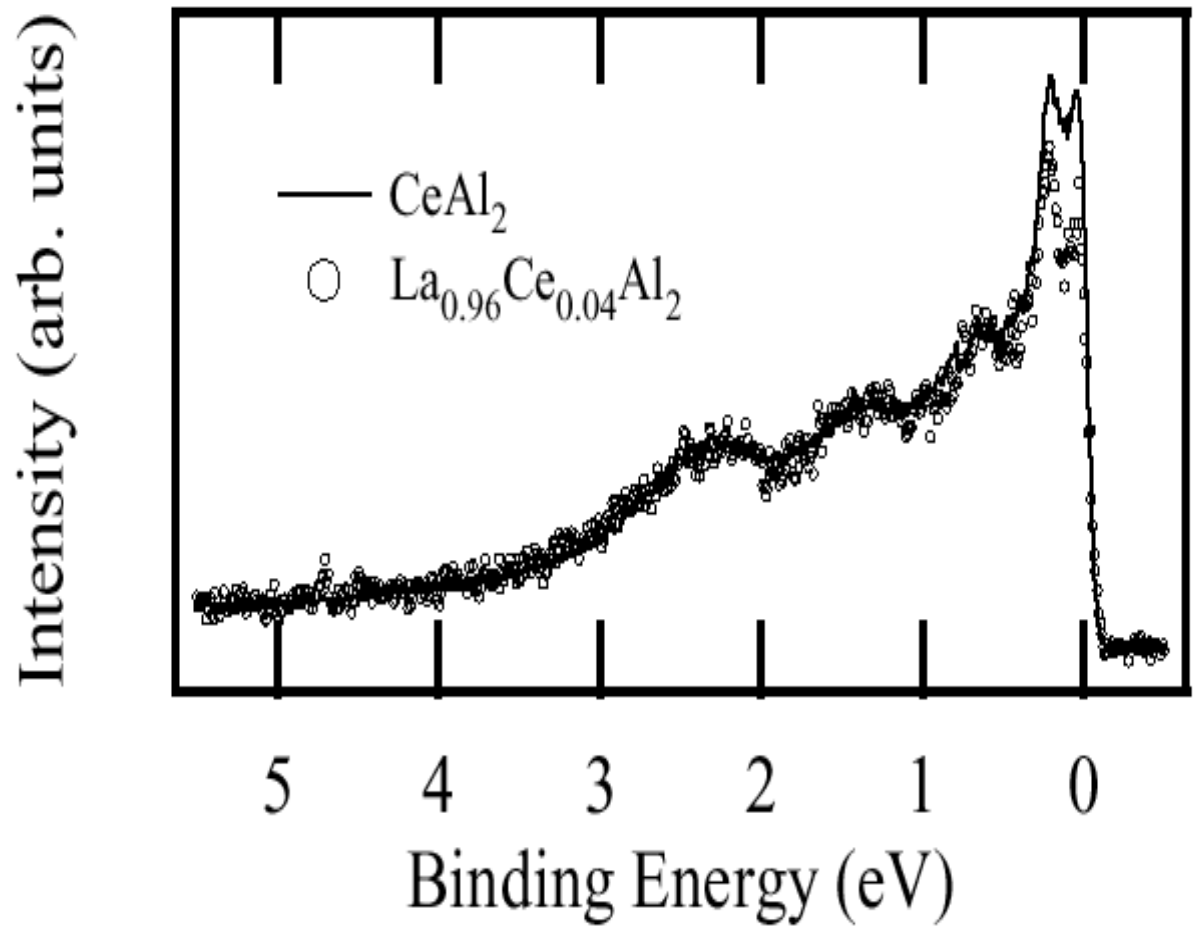

Fig. 3

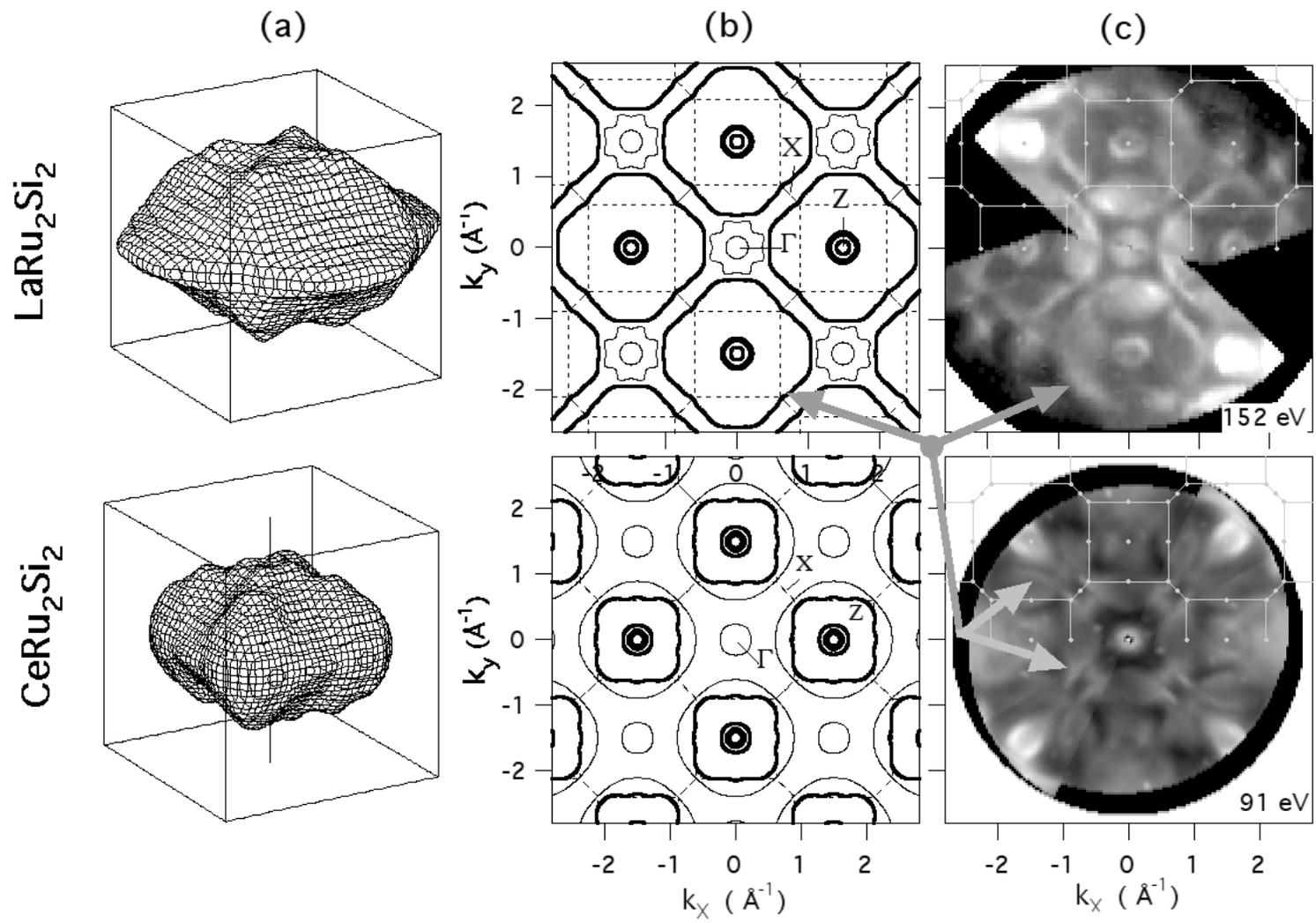

Fig. 4

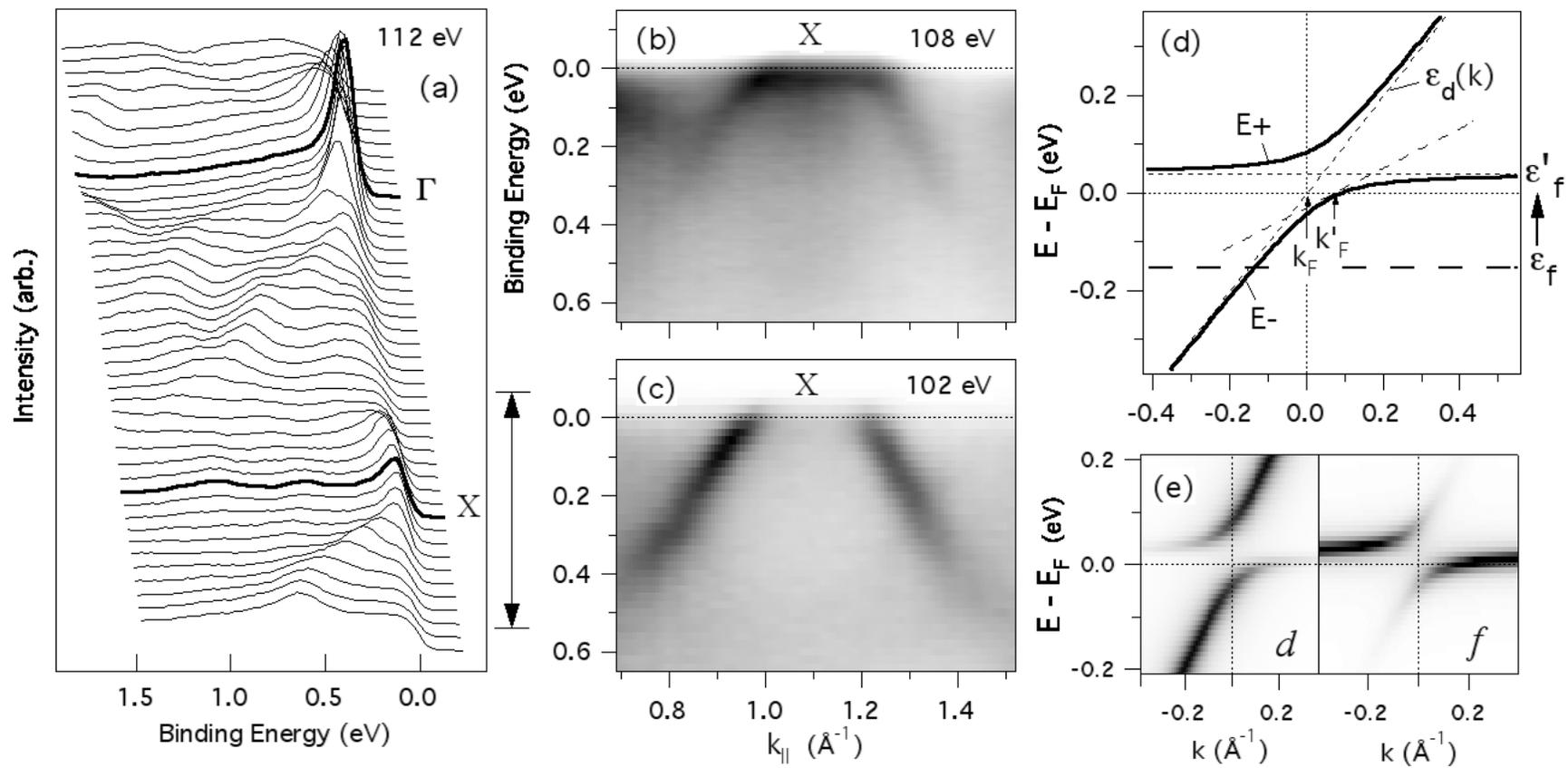

Fig. 5

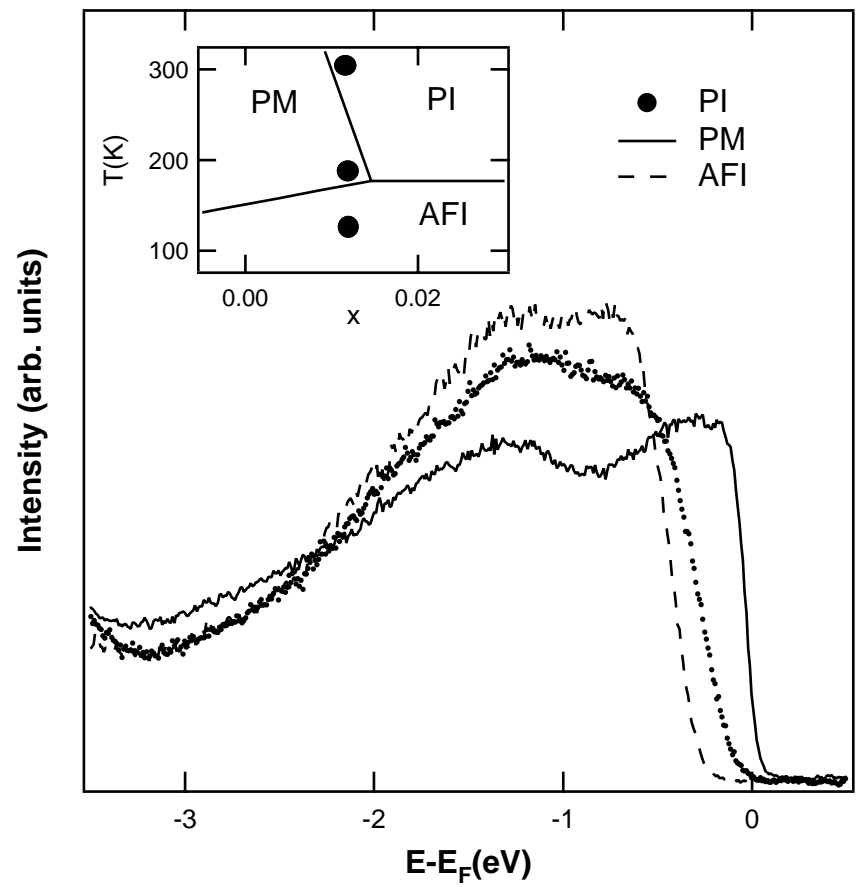

**Fig. 6**

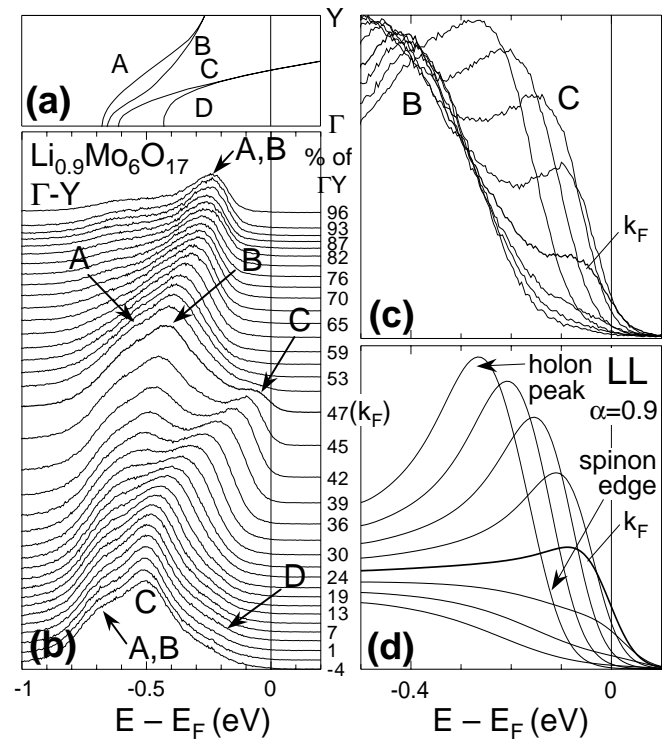

**Fig. 7**

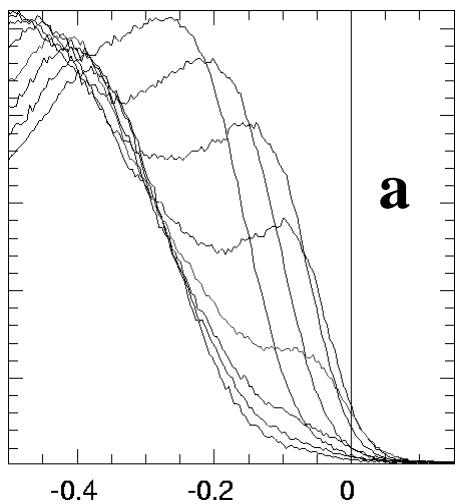
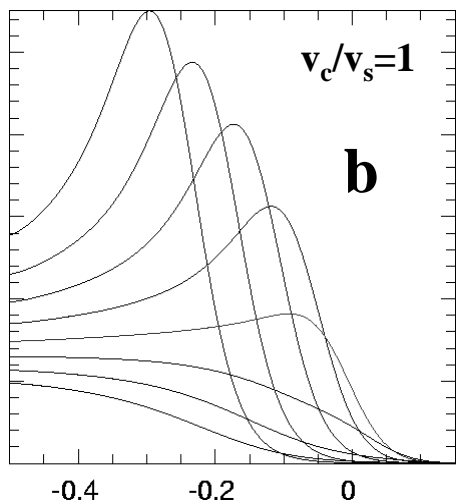
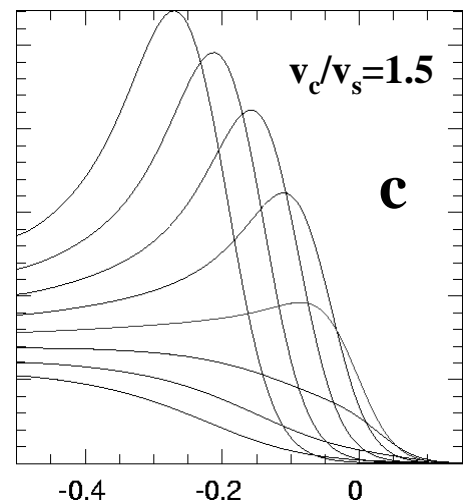
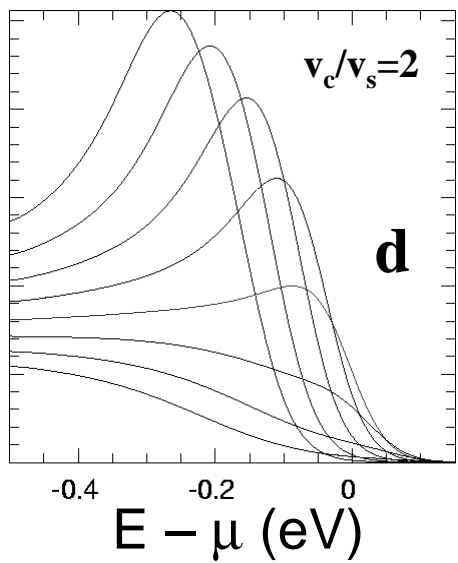
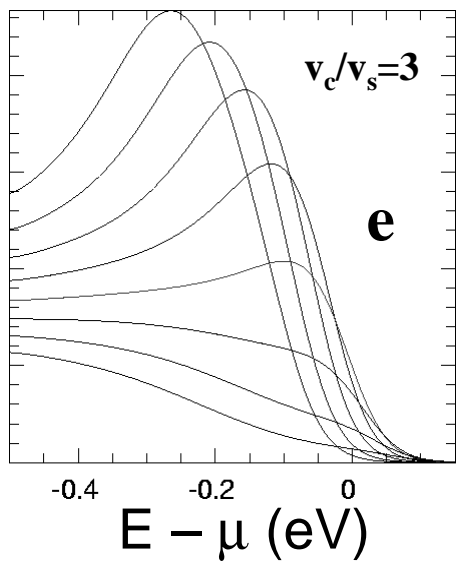
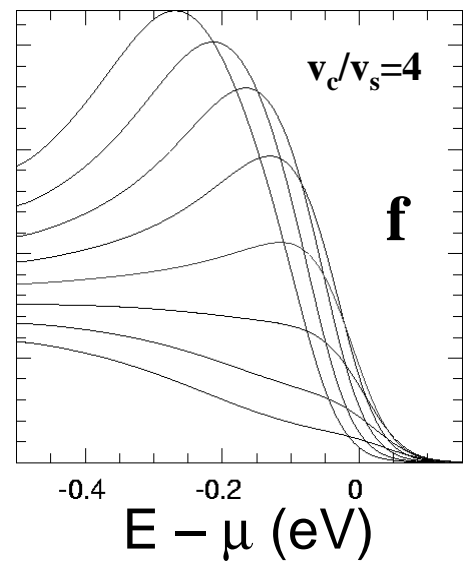

**Fig. 8**

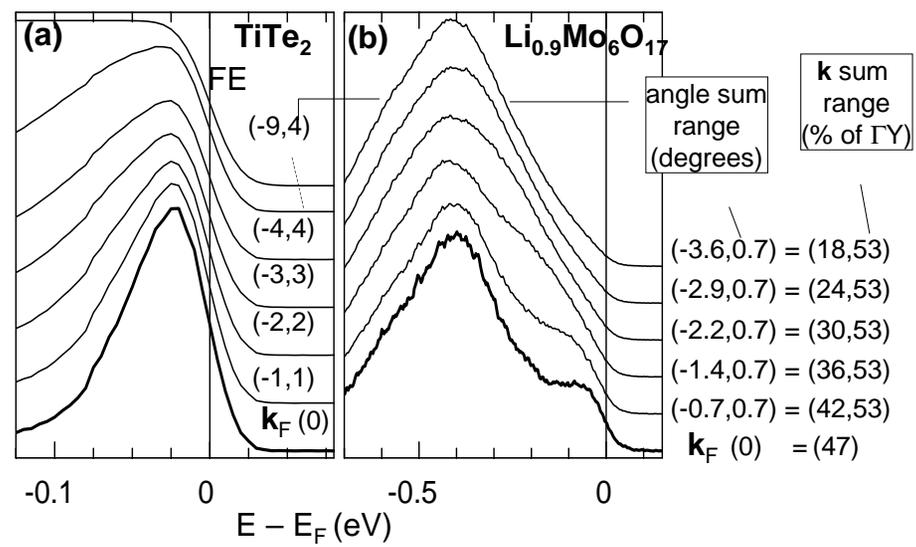

**Fig. 9**